\documentclass[a4paper,twoside]{article}
\newcommand{\affil}[1]{$^{\rm #1}$}
\usepackage[authoryear]{natbib}
\bibpunct{(}{)}{;}{a}{}{,}
\usepackage{graphicx}
\date{} 
\title{\large\bf\flushleft Photometric Observations of the $\eta$ Carinae 2009.0 Spectroscopic Event}
\author{\parbox{\textwidth}{\flushleft
\vspace{-0.5cm}
{\it H. Landes\affil{A,B} and M. Fitzgerald\affil{A}}\\
\vspace{0.4cm}
{\small \affil{A}\,School of Physics, Monash University, VIC, 3800 Australia}\\
{\small \affil{B}\,Email: herschel.landes@sci.monash.edu.au}}}%
\begin{document}
\twocolumn[
\begin{changemargin}{.8cm}{.5cm}
\begin{minipage}{.9\textwidth}
\vspace{-1cm}
\maketitle
%
%
\small{\bf Abstract:}
We have observed $\eta$ Carinae over 34 nights between 4th January 2009 and 27th March 2009 covering the estimated timeframe for a predicted spectroscopic event related to a suspected binary system concealed within the homunculus nebula. A photometric minimum feature was confirmed to be periodic and comparison to a previous event indicated that the period to within our error at 2022.6$\pm1.0$ d. Using the E-region standard star system, the apparent $V$ magnitudes determined for the local comparison stars were HD303308 8.14$\pm0.02$, HD 93205 7.77$\pm0.03$ and HD93162 8.22$\pm0.05$. The latter star was found to be dimmer than previously reported.


\medskip
\medskip
\end{minipage}
\end{changemargin}
]
\small

\section{Introduction}
$\eta$ Carinae is a very well studied variable stellar system classified as a ‘Luminous Blue Variable’ or S Dor star although with some unique features (van Genderen et al 1999). Over the last two centuries, $\eta$ Carinae has gone through changes in its brightness over a range of 9 magnitudes. The historic light curve from the 16th century (see Figure 1) to the present identifies the great eruption in 1843 brightening the $\eta$ Carinae system to -1 magnitude at its peak, with a heavy drop off to a low of 8th magnitude around 1900, and then a slow brightening at a rate of approximately 0.05 magnitudes per year (van Genderen et al. 1994, van Genderen et al. 1999, Frew 2004, Fernandez-Lajus et al. 2009.)
\begin{figure}[h]
\begin{center}
\includegraphics[scale=0.5, angle=0]{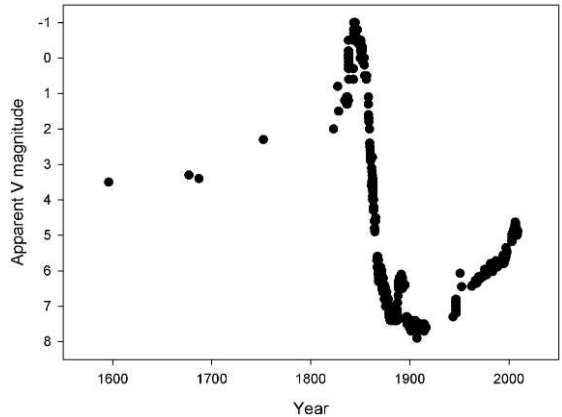}
\caption{Historic light curve of $\eta$ Carinae based on magnitudes published by van Genderen et al. 1994, van Genderen et al. 1999, Frew 2004, Fernandez-Lajus et al. 2009.}\label{figexample}
\end{center}
\end{figure}
Feinstein \& Marraco (1974) speculated on the existence of a possible periodic variation of flux in $\eta$ Carinae of period approximately 1100 d while Antokhin \& Cherepashchuk (1993) identified a possible model for $\eta$ Carinae as a close binary system at the common-envelope stage, but found no evidence of periodicity.

Damineli et al. (1996) was the first to suggest a 5.52 year binary-related cycle. This result was tested with a multi-wavelength observation program covering the 1998.0 event. The eclipse lasted three months, and the variation of $\eta$ Carinae at radio wavelengths, in its spectra in the near IR and at optical wavelengths, and in its X-ray brightness led to a binary model for the system (see Corcoran 2002, and references therein). The suggested orbit has $e$ $\approx$ 0.9 and component masses of ~100$M_\odot$ and 30-60$M_\odot$ with large uncertainty in each parameter.

A further multi-wavelength observation program covering the next (2003.5) event has seemingly dispelled the doubts of a binary companion and concluded the average period, over 12 different spectral bands, to be  2022.7 $\pm$ 1.3 d (Damineli et al. 2008). In this paper, Damineli questioned the long-term stability of the period given the high mass loss rate of the companion stars and possible tidal interaction during periastron passage.
\section{Observations}

In an endeavour to test the periodic nature of a photometric minimum feature that coincides with the periastron event, $\eta$ Carinae was observed with the 12$"$ Monash Automated Observatory (MAO) over 34 nights between  4th  January 2009 and 27th  March 2009, covering the estimated range of the predicted event. Images were recorded with an SBIG ST7 camera (715x510 pixels) over a 12$^{\prime}$ by 8$^{\prime}$ field of view, with 2x2 binning (1.9$^{\prime\prime}$/pixel scale) and through a filter closely approximating the standard Johnson $V$. Although $B$ and $R$ filters were available, the $B$ filter was not used due to its significantly different characteristic from the standard Johnson $B$ filter, while the $\eta$ Carinae system was too bright for sufficiently long exposures in R due to its large $H\alpha$ emission. Photometry in the $V$ band is relatively free of strong emission lines but many authors have described difficulties in transforming from instrument magnitudes to a standard system (eg. Sterken et al. 1999, Martin et al 2004 ).

\begin{figure}[h]
\begin{center}
\includegraphics[scale=0.75, angle=0]{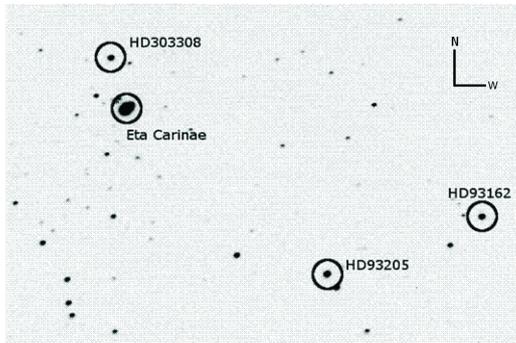}
\caption{The $\eta$ Carinae field, 2x2 binned, 1.9$^{\prime\prime}$ pixel scale image showing the comparison star HD 303308 and check stars HD 93162 and HD 93205. Stars are shown with sample apertures. Note the inclusion of two very dim stars, Tr16-64 and Tr16-65, within the aperture for $\eta$ Carinae.}\label{fig2}
\end{center}
\end{figure}

Aperture photometry was performed on the images by the use of a C program specifically designed to process images from the MAO telescope and detector system.  The centroid was determined as the brightest pixel within the aperture. While more accuracy may have been possible by determining the centroid from the flux distribution, this method was found to perform quite suitably for the size of apertures used. The aperture radius was determined by the minimum radius at which the difference of instrumental magnitudes between one radius and the next was minimal. This was estimated to be 9 pixels. Increasing the aperture radius increases the noise, while decreasing the aperture radius would result in significant reductions in the flux being measured from the extended homunculus of $\eta$ Carinae. The sky is defined as background contribution within the aperture. This was found by determining the average pixel count over all pixels where the counts were less than three standard deviations from the median count for the entire frame. An image from the MAO with sample apertures is provided in Figure 2. The stars Tr16-65 ($V$=11.09) and Tr16-64 ($V$=10.72) (Feinstein et al. 1973) adjacent to $\eta$ Carinae, are unavoidably included in the aperture. They are assumed to be stable and their flux contribution negligible.

The flux recorded from $\eta$ Carinae includes both the central stellar sources (primary and the secondary) and the homunculus nebula surrounding them. Ground-based telescopes are unable to resolve the central source from the homunculus. While it is possible that the homunculus could detrimentally affect our photometric observations, it has been shown that ground-based photometry is still able to resolve brightness variations, potentially at different amplitudes due to temporal smearing (Martin et al. 2004), although it has been argued that such smearing is excluded (van Genderen \& Sterken 2004)

Minor Planet Observer software (Brian Warner) was used to obtain bias and dark frame corrected images, with new bias and dark frames being taken on average every 10 minutes. Images were later flat-fielded using twilight sky flats by the standard method using IMCOMBINE and CCDPROC within IRAF. One combined master flat-field image with mean counts at approximately 19,000 ADU, was used for all observations.

As there was concern that there may have been some variation in the comparison star magnitude, it was decided to measure the $V$-band magnitude of the comparison and check stars. Magnitudes in the MAO $V$ system were achieved by monitoring the E region standards, e428, e439, e478, e528, e537, e570 and e633. (Menzies et al. 1989). A simultaneous chi-squared fit to the photometric solution, comprising the first order extinction and $V-R$ colour terms, was obtained from observations of the standard stars on various nights. The $V$ magnitude of the comparison star was determined to be 8.14 $\pm$ 0.02 for HD303308. This result is comparable to the range of values previously reported for this star over the last 30 years: 8.17 (Feinstein, Marraco \& Mirabel 1973), 8.15 (Feinstein 1982), 8.19 (Massey \& Johnson 1993), 8.12 (Antokhin \& Cherepashchuk 1993) and 8.13 mag (ESA 1997). It seems possible that this star varies over the longer term or is perhaps sensitive to different photometric systems.

To test the stability of the comparison star, HD 303308, the differential magnitude with comparison to HD 93205 was monitored. Over the total period of observations, the differential magnitude between the check and comparison stars were found to be stable at $-$0.36 ±$\pm$ 0.01 magnitudes (see Figure 3).

HD 93205 (check star 1) was determined to have an apparent magnitude of 7.77 $\pm$ 0.03, with previously stated magnitudes of 7.76 (Feinstein 1982) and 7.75 (Feinstein, Marraco \& Mirabel 1973). Feinstein, Marraco \& Mirabel (1973). HD 93162 (check star 2) was determined to have an apparent magnitude of 8.22 $\pm$ 0.05, while previously it was stated to be 8.10 (Feinstein, Marraco \& Mirabel 1973) and 8.09 (Feinstein 1982) magnitudes, showing that it has dimmed significantly over the last few decades. Published magnitudes for this star were stable at 8.10 until the late nineties when deviations began : 8.11 (Massey \& Johnson 1993), 8.13 (Morrison et al. 2001) and 8.16 (ESA 1997)

\section{Results}

Differential magnitudes for $\eta$ Carinae were calculated for each frame relative to the comparison star HD 303308. We attempted to minimise the overall uncertainty in our measurements by taking as many images over the night as possible. Over the course of an observation night, between 12 and 365 images were obtained with a median number of 47 per night, the number largely depending on the weather conditions. Wherever possible, only images taken for $\sec z$ less than 1.2 were used. The median differential magnitude of these images was determined and the uncertainty estimated as the standard deviation of the measured magnitudes over each observing session. Any results greater than 2$\sigma$ from the median value of differential magnitudes for a given session were ignored. The apparent magnitude of $\eta$ Carinae was estimated from the measured apparent magnitude of HD 303308 ($V$=8.14). The standard deviation (SD) of the night’s observations is given. Due to the large number of photons collected from Eta Carinae itself, the standard devation of the observations result mainly from the dimmer comparison stars used. This also led to the larger standard deviations in the comp-check observations than in the Eta Carinae observations. Results are tabulated in Table 1 and shown in graphical form in Figure 3.
\begin{table}[h]
\begin{center}
\caption{Observations. }\label{tableexample}
\begin{tabular}{ |c|c|c|c|c|}
\hline JD$^a$&App Mag&SD&N \\
\hline 4836.162&4.922&0.019&27 \\
 4839.154&4.932&0.015&35 \\
 4842.172&4.917&0.021&53 \\
 4843.144&4.911&0.024&47 \\
 4844.154&4.891&0.012&29 \\
 4845.147&4.890&0.015&24 \\
 4849.183&4.936&0.014&37 \\
 4850.152&4.957&0.012&35 \\
 4851.080&4.963&0.010&100 \\
 4851.966&4.980&0.017&15 \\
 4852.901&5.017&0.023&51 \\
 4856.144&5.025&0.013&92 \\
 4859.158&5.040&0.020&153 \\
 4860.116&5.049&0.019&47 \\
 4866.986&5.014&0.021&32 \\
 4867.973&5.021&0.017&12 \\
 4869.106&5.014&0.017&27 \\
 4877.099&4.993&0.018&365 \\
 4878.102&4.997&0.027&105 \\
 4879.088&4.980&0.018&232 \\
 4880.041&4.995&0.021&71 \\
 4883.990&4.949&0.014&14 \\
 4885.003&4.952&0.019&52 \\
 4886.941&4.946&0.014&41 \\
 4888.057&4.927&0.028&115 \\
 4899.050&4.886&0.017&101 \\
 4900.048&4.879&0.017&119 \\
 4903.066&4.889&0.023&74 \\
 4907.956&4.897&0.019&39 \\
 4909.979&4.897&0.019&33 \\
 4910.985&4.890&0.016&66 \\
 4913.950&4.898&0.016&41 \\
 4916.939&4.896&0.014&58 \\
 4917.917&4.885&0.021&16 \\
\hline
\end{tabular}
\medskip\\
$^a$ JD-2450000\\
From left: Central Julian Date for a  night’s observations, apparent magnitude (Diff-Mag + 8.14), Standard Deviation of nights measurements, Number of observations used.
\end{center}
\end{table}

\begin{figure}[h]
\begin{center}
\includegraphics[scale=0.45, angle=0]{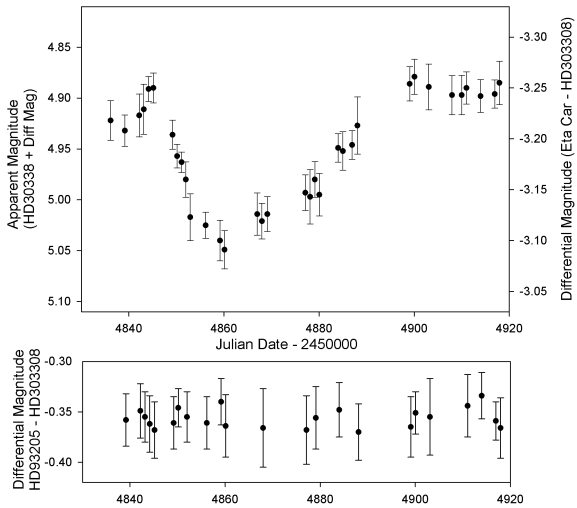}
\caption{Top: Differential magnitude and apparent magnitude of $\eta$ Carinae, derived as in the text, versus JD. Bottom: Differential magnitude between comparison and check star, showing stability. Within both plots, error bars represent the sample standard deviation.}\label{fig3}
\end{center}
\end{figure}

The spectroscopic event is defined by phase 0 being the time of minimum excitation of HeI $\lambda$6678 emission lines (Daminelli et al 2008), with an ephemeris JD=2452819.8+2022.8$E$ (Fernandez-Lajus et al. 2009) This collapse is assumed to coincide with periastron passage (Damineli et al. 2000). Close to periastron passage, accretion of the primary wind onto the companion takes place (Soker 2005) and a thick belt is formed, blocking the ionizing radiation of the secondary in equatorial directions (Kashi \& Soker 2009)

Approximately one day after the predicted collapse of HeI emission lines on Janurary 11th 2009, we observed a peak in brightness that preceded the minimum event featured on the light curve. Comparison with the $V$ band light curve observed by (Fernandez-Lajus et al. 2009) in 2003.5 shows a similar asymmetric pattern, with a distinct maximum before the initial decline towards the minima, a comparison between these two lightcurces is shown in Figure 4. This well defined minimum point measures 0.16 mag in depth for the 2009.0 event compared to 0.11 mag for the 2003.5 event. The timescale of the decline from initial maximum to minimum seems to have remained constant ($dt=15$ d) between the 2003.5 and 2009.0 event while earlier observations in the infrared bands for the 1998.0 event also suggest a similar timescale, albeit with less complete data (Feast et al. 2001). 

A final maximum is seen in our data some 39 days after the minimum, however the 2003.5 data, due to incompleteness, does not identify this maximum point, although it does suggest that the final maximum point does not occur at the same phase location, perhaps implying that the timescale from minimum to final maximum is not a stable function of orbital phase. The  minimum  feature having the same period as the spectroscopic event  and similar asymmetry confirms the observed photometric minimum is part of the periastron passage phenomena. The variable extent and duration of the minimum featured compared with previous observations indicates a dynamic process is occurring. 

The 15d time interval between the maxima and minima was compared for the 2003.5 and 2009.0 data by selecting the period which gave the smallest combined deviation between the determined period from the minima and initial maximum points on the light curve. This value was measured to be 2022.6$\pm1.0$ d, where the error was estimated from the size of the deviation from a perfect fit, which could not conceivable be smaller than a day due to the nature of the observations taken.

The maximum-minimum time interval, having a similar period as the spectroscopic event, confirms this photometric feature to be part of the periastron passage phenomena. The consistency of this 15 d maximum to minimum time interval for the 2009.0 and 2003.5 events may be coincidental, but it may be possible to also speculate a link to the orbital parameters of the secondary star. Alternatively, the variable depth of the minimum and duration of the second part of the event could indicate that these are dynamic features relating to possible accretion, cooling and dispersion of material over the passage event.
\begin{figure}[h]
\begin{center}
\includegraphics[scale=0.5, angle=0]{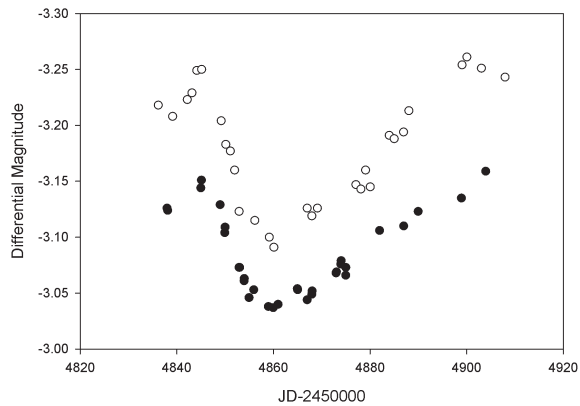}
\caption{Comparison of the 2009.0 $\eta$ Carinae periastron event (Empty circles) to the 2003.5 event (Filled circles) (Fernandez-Lajus et al. 2009). The 2003.5 data has been shifted forward 2022.5d, but the differential magnitude has been left unchanged.}\label{figexample}
\end{center}
\end{figure}
\section{Conclusion}

Within the limits of ground-based observations, we conclude the periodicity of the asymmetric minimum feature in the light curve and that it  is linked to the periastron passage of the Eta Carinae binary companion. We determined this minimum to have a period 2022.6$\pm1.0$ d. We have observed the coincidence of the 15 day time interval between maximum and minima but that the depth of the minima and duration have varied in comparison with the last event indicating the dynamic nature of this feature. We have also observed that HD 93162 appears to be significantly dimmer ($V$=8.22$\pm$0.05) than previously reported.

\section*{Acknowledgments} 

IRAF is distributed by the National Optical Astronomy Observatories, which are operated by the association of Universities for Research   in Astronomy, Inc., under cooperative agreement with the National  Science Foundation. The guidance of Don Whiteman and Barry Gerdes of Bintel was invaluable in addressing certain technical difficulties with the Meade Telescope as was Denis Coates' and Amit Kashi's helpful insights into the final manuscript.

\section*{References} 

\flushleft

Antokhin, I.I., Cherepashchuk, A.M., 1993, ARep, 37,152

Corcoran M. F., 2002, X-ray Variability of Eta Carinae. Eta Carinae: Reading the Legend, Conference, Mount Rainier, WA, USA, 11-12 July 2002.

Damineli, A., 1996, ApJ, 460, L49

Damineli, A., Kaufer, A., Wolf, B. et al., 2000, ApJ, 528, L101

Damineli, A., Hillier, D., Corcoran, M.F., et al., (2008), 384, 1649

ESA SP-1200, 1997, The Hipparcos and Tycho Catalogues. ESA Publications Division, Noordwijk

Feast, M., Whitelock, P., Marang, F., 2001, MNRAS, 322, 741

Feinstein, A., 1982, AJ, 87, 1012

Feinstein, A., Marraco, H.G., Mirabel, I. 1973, A\&AS, 9, 233

Feinstein, A., Marraco, H.G., 1974, A\&AS, 30, 271

Feinstein, A., Marraco, H.G., Muzzio, J.C., 1973 A\&AS, 12, 331

Fernandez-Lajus, E.,  Farina, C., Torres, A.F., et al., 2009, A\&A, 493, 1093

Frew, D.J., 2004, JAD, 10,6

Kashi A., Soker N., 2009, NewA, 14, 11

Harris, A.W., Young, J.W., Bowell, E., et al., 1989, Icar, 77,171

Laing, J.D., 1989, SAAOC, 13, 1

Martin, J.C., Koppelman, M.D., The HST $\eta$ Carinae Treasury Project Team, 2004, AJ, 127, 2352

Massey, P., Johnson, J., 1993, AJ, 105, 980

Menzies, J.W, Cousins, A.W.J., Banfield, R.M., Laing, J.D., 1989, SAAOC, 13, 1

Morrison, J. E., Röser, S., McLean, B.,et al., 2001 AJ, 121, 1752

Soker, N. 2005,  ApJ, 635, 540

Sterken, C., Freyhammer, L., Arentoft, T., van Genderen, A.M., 1999, A\&A 346, 33S

van Genderen, A.M., de Groot, M.J.H, The, P.S., 1994, A\&A, 283, 89

van Genderen, A.M., Sterken, C., de Groot, M., Burki, G., 1999, A\&AS, 343, 847

van Genderen, A.M., Sterken, C., 2004, A\&A, 423, L1

van Genderen, A.M., Sterken. C., Allen, W.H., Walker, W.S.G., 2006, JAD, 12, 3


\end{document}